\documentclass[9pt,twocolumn,twoside]{opticajnl}
\journal{opticajournal} 

\setboolean{shortarticle}{false}

\title{Subnatural-linewidth fluorescent single photons.}

\author[1,2]{He-bin Zhang}
\author[2,4]{Gao-xiang Li}
\author[1,3,5]{Yong-Chun Liu}

\affil[1]{State Key Laboratory of Low-Dimensional Quantum Physics, Department of Physics, Tsinghua University, Beijing 100084, People’s Republic of China}
\affil[2]{Department of Physics, Huazhong Normal University, Wuhan 430079, China}
\affil[3]{Frontier Science Center for Quantum Information, Beijing 100084, People’s Republic of China}

\affil[4]{e-mail: gaox@mail.ccnu.edu.cn}
\affil[5]{ycliu@tsinghua.edu.cn}

\begin{abstract} 
Subnatural-linewidth single photons are of vital importance in quantum optics and quantum information science.
According to previous research, it appears difficult to utilize resonance fluorescence to generate single photons with subnatural linewidth.
Here we propose a universally applicable approach for generating fluorescent single photons with subnatural linewidth, which can be implemented based on $\Lambda$-shape and similar energy structures. Further, the general condition for obtaining fluorescent single photons with subnatural linewidth is revealed. The single-photon linewidth can be easily manipulated over a broad range by external fields, which can be several orders of magnitude smaller than the natural linewidth. 
Our study can be easily implemented in various physical platforms with current experimental techniques, and will significantly facilitate the research on the quantum nature of resonance fluorescence and the technologies in quantum information science.
\end{abstract}

\begin{document}
	
\maketitle

\section{Introduction}
\label{Sec-1}
Single-photon character of light means a special antibunching quantum nature indicated by the normalized second-order correlation function $g^{(2)}(0)=0$, which is of crucial importance in quantum communication~\cite{Bennett1984_Quantum, Bennett1992_Experimental} and optical quantum computing~\cite{Knill2001_A-scheme, O'Brien2009_Photonic}. It is a requirement for single-photon source, and may be required for single modes of multiphoton source, e.g., entangled two-photon source. Fortunately, the simple resonance fluorescence from single emitter, including atom~\cite{Kimble1977_Photon, Grangier1986_Observation} and various atom-like systems~\cite{Michler2000_A-Quantum, Kurtsiefer2000_Stable, Lounis2000_Single}, satisfies perfect single-photon character. Naturally, fluorescent photons have always been a fundamental element in the fields of quantum optics, photonics and quantum information science. 

The conversion of quantum state between flying photonic and stationary qubits is the basic physical process for quantum information storage and processing, and is therefore crucial for long-distance quantum communication~\cite{Duan2001_Long} and quantum networks~\cite{Yao2005_Theory, Kimble2008_The-quantum}. To achieve efficient interfaces between photons and atoms, the photons are required to have a bandwidth smaller than the linewidth of atomic transitions~\cite{Liao2014_Subnatural-Linewidth, Horiuchi2014_Subnatural-linewidth, Shu2016_Subnatural-linewidth, Matthiesen2012_Subnatural}.  
In recent years, research on subnatural-linewidth photons based on atom and quantum dot has received tremendous attention and gained significant progress~\cite{Matthiesen2012_Subnatural, Nguyen2011_Ultra-coherent, Yang2018_Tomography, Du2008_Subnatural, Liao2014_Subnatural-Linewidth, Shu2016_Subnatural-linewidth, Du2008_Nonlinear}.

However, it is recently demonstrated in Refs.~\cite{Hanschke2020_Origin, Carreno2018_Joint, Phillips2020_Photon} that for the resonance fluorescence from a two-level system, single-photon character and subnatural linewidth cannot be satisfied simultaneously, which means that the fluorescent single photons cannot have subnatural linewidth. These studies reveal a previously neglected point that the individual spectral analyzer or interferometer can only measure the properties of main fluorescent components, but cannot faithfully reflect the nature of fluorescent single photons. Alternatively, the measurement scheme of adding a filter in front of the Hanbury-Brown Twiss (HBT) setup can be used to estimate the linewidth of single photons. According to Refs.~\cite{Hanschke2020_Origin, Carreno2018_Joint, Phillips2020_Photon}, the perfect single-photon character of fluorescence can be accounted for the destructive interference between the coherent and incoherent components. It is demonstrated that the spectral filtering with a bandwidth approaching the natural linewidth of emitter can significantly spoil the single-photon character of fluorescence due to the partial loss in the incoherent component. When most fluorescent components are focused on an ultranarrow coherent peak~\cite{Matthiesen2012_Subnatural, Nguyen2011_Ultra-coherent}, the above conclusions remain valid~\cite{Hanschke2020_Origin, Carreno2018_Joint, Phillips2020_Photon}, which can be demonstrated by the destruction of single-photon response on the detector. Consequently, the linewidth of fluorescent single photons appears to be limited by the natural linewidth of emitter. 
A natural question is whether it is possible to generate subnatural-linewidth single photons based on resonance fluorescence. 

Here we propose a universally applicable approach for generating single photons with subnatural linewidth utilizing atom or atom-like system. Specifically, we introduce a quantum emitter with a long-period transition loop as shown in Fig.~\ref{Fig-1}(a). We find that all the spectral components of fluorescence radiated by the emitter can be focused on a bandwidth that is smaller than the natural linewidth. Therefore, the subnatural-linewidth single photons are achieved. Besides, the linewidth of single photons can be easily manipulated over a broad range by varying the intensities of external fields. Thereby, the single-photon linewidth several orders of magnitude smaller than the natural linewidth can also be readily obtained. Worth mentioning is that we also reveal the general condition for generating single photons with subnatural linewidth utilizing atom or atom-like system, i.e., the successive emission process of the target single photons is totally dominated by the transition loop with a metastable state. This condition can be satisfied in various physical systems with $\Lambda$-shape and similar energy structures, which exhibits the high experimental feasibility of our proposed approach.

\section{Subnatural-linewidth single photons}
\label{Sec-2}

\subsection{Model of emitter and detector}

\begin{figure}[ht!]
	\centering\includegraphics[draft=false, width=0.9\columnwidth]{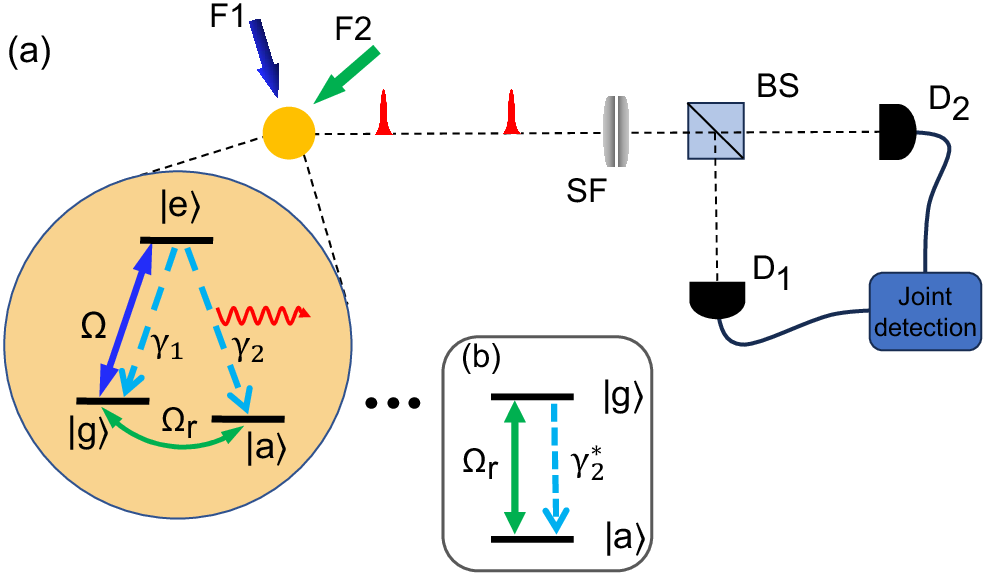}
	\caption{(a) Schematic diagram of the emitter and the detection setup. On the left is a $\Lambda$-shape emitter including ground states $|g\rangle$ and $|a\rangle$, and excited state $|e\rangle$. The transitions $|g\rangle\leftrightarrow|e\rangle$ and $|g\rangle\leftrightarrow|a\rangle$ are driven by the external fields F1 and F2, respectively. Fluorescent photons emitted from the transition $|e\rangle\rightarrow |a\rangle$ are collected by a  HBT setup. Moreover, a spectral filter (SF) is inserted between the emitter and the HBT setup, and thus the filtering frequency and bandwidth determine the detection frequency and bandwidth, respectively. (b) Schematic diagram of the effective level system of the emitter in (a) in the weak excitation regime.\label{Fig-1}}
\end{figure}

As shown in Fig.~\ref{Fig-1}(a), a closed $\Lambda$-shape system is considered as the quantum emitter, where the transition $|g\rangle\leftrightarrow|e\rangle$ is driven by a laser field F1, and the transition $|g\rangle\leftrightarrow|a\rangle$ is driven by another coherent field F2. A transition loop composed of the transitions $|g\rangle\rightarrow|e\rangle \xrightarrow{\rm{emission}} |a\rangle$ and $|a\rangle\rightarrow|g\rangle$ is constructed in the emitter by these two external fields.
The fluorescence from the transition $|e\rangle\rightarrow |a\rangle$ is collected by the detector, which comprises a standard HBT setup and a spectral filter inserted between the emitter and the HBT setup. The detector is used to test whether subnatural-linewidth single photons are generated~\cite{Hanschke2020_Origin, Phillips2020_Photon}. 
In the rotating frame at the driving frequencies, the Hamiltonian of the system consisting of emitter and detector shown in Fig.~\ref{Fig-1}(a) is given by 
\begin{equation}
	\textit{H}=\Delta_e\sigma_{ee} + \Delta_a\sigma_{aa} + \Delta_s s^{\dag}s + (\Omega \sigma_{eg} + \Omega_r \sigma_{ga} + g \sigma_{ea}s + \rm{H.c} ).
	\label{Eq-1}
\end{equation}
Here the detector is modeled by a quantized harmonic oscillator with bosonic annihilation operator $s$. $\Delta_e$ and $\Delta_a$ ($\Delta_s$) represent the frequency detunings between the transitions $|g\rangle\leftrightarrow |e\rangle$ and $|g\rangle\leftrightarrow |a\rangle$ (detector) and the external fields. Without loss of generality, we set $\Delta_e=\Delta_a=\Delta_s=0$ for simplicity. $\Omega$ and $\Omega_r$ respectively represent the Rabi frequencies of the external fields F1 and F2.
$g$ is the coupling coefficient between the detector and the detected transition $|e\rangle\rightarrow |a\rangle$.
Therefore, the evolution of the combined system composed of the emitter and detector is governed by the master equation
\begin{equation}
	\dot{\rho}=-i [\textit{H},\rho] + \gamma_{1} \mathcal{D}[\sigma_{ge}]\rho + \gamma_{2} \mathcal{D}[\sigma_{ae}]\rho + \kappa D[s]\rho,
	\label{Eq-2}
\end{equation}
where $\mathcal{D}[o]\rho\equiv o\rho o^{\dag} - \frac{1}{2}\rho o^{\dag}o - \frac{1}{2}o^{\dag}o \rho$ is the Lindblad superoperator. $\gamma_{1}$ and $\gamma_{2}$ represent the natural linewidths of the decays $|e\rangle\rightarrow |g\rangle$ and $|e\rangle\rightarrow |a\rangle$, respectively.
The detection bandwidth $\kappa$ and its inverse, respectively, reflect the frequency and time resolutions of detector, therefore, the constraints imposed by the uncertainty principle on the detection process are included in a self-consistent way.

\subsection{Effective level system of emitter.}
\label{Sec-2b}

In the limit of the vanishing coupling between the emitter and detector, i.e., $g\rightarrow0$, the detector can be regarded as a passive object~\cite{Carreno2018_Joint, Valle2012_Theory}.
Therefore, it is beneficial to study the nature of the emitter before considering the detection response for this quantum source.
In the weak excitation regime of the emitter, i.e., $\Omega, \Omega_r\ll\gamma_1,\gamma_2$, the evolution rate of the excited state is much greater than that of the ground states.
Therefore, the excited state can be adiabatically eliminated, which induces effective decays between the ground states, i.e., 
\begin{equation}
	\mathcal{L}_{\rm{eff}}\rho_g=\gamma^{*}_{1} \mathcal{D}[\sigma_{gg}]\rho_g + \gamma^{*}_{2} \mathcal{D}[\sigma_{ag}]\rho_g.
	\label{Eq-3}
\end{equation}
The first term denotes the dephasing of the state $|g\rangle$, and the second term denotes the effective decay $|g\rangle\rightarrow|a\rangle$ (see Appendix~\ref{App-1} for details).
The effective linewidths satisfy 
\begin{equation}
\gamma^{*}_{i}= \dfrac{4\gamma_i|\Omega|^2}{(\gamma_1+\gamma_2)^2},
\label{Eq-3a}
\end{equation}
with $\textit{i}=1,2$, which can be reduced to $\gamma^{*}=|\Omega|^2/\gamma$ with $\gamma_1=\gamma_2=\gamma$.

Therefore, the emitter in Fig.~\ref{Fig-1}(a) can be regarded as an effective two-level system driven by a coherent field as shown in Fig.~\ref{Fig-1}(b).
The effective decay $|g\rangle\rightarrow|a\rangle$ corresponds to a two-photon process in the original level system, where the emitter absorbs the first photon by the transition $|g\rangle\rightarrow|e\rangle$ driven by the laser, and subsequently emits the second photon by the spontaneous decay $|e\rangle\rightarrow|a\rangle$.
Remarkably, different from the normal two-level system, the linewidth of the effective two-level system is determined by the ratio of the laser intensity to the emitter's natural linewidth and thus can be set artificially.

\subsection{Spectral Property of Fluorescent Photons.}

\begin{figure}
\centering\includegraphics[draft=false, width=1\columnwidth]{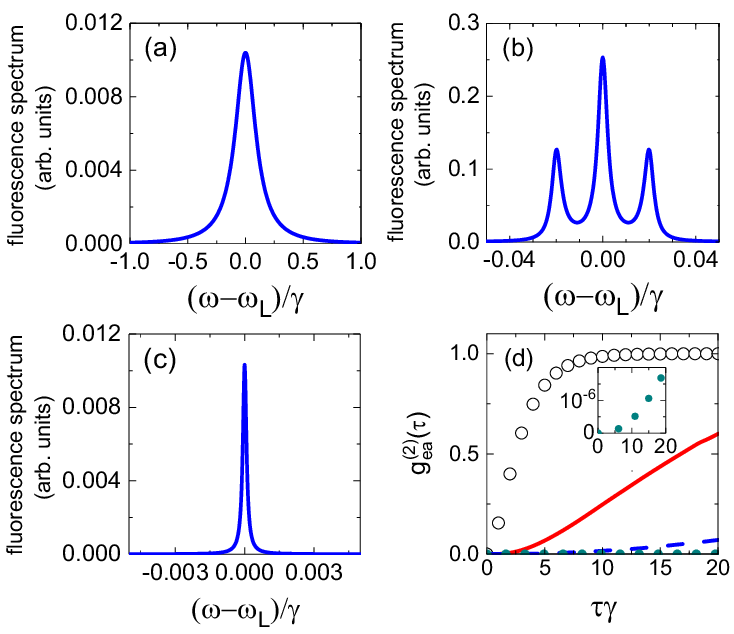}
\caption{Fluorescence spectrum (incoherent component) for the transition $|e\rangle\leftrightarrow|a\rangle$ for  $\Omega =1/\sqrt{10}\gamma$, $\Omega_r=10^{-2}\gamma$ in (a), $\Omega =5\times10^{-2}\gamma$, $\Omega_r=10^{-2}\gamma$ in (b), and $\Omega =10^{-2}\gamma$, $\Omega_r= 10^{-5}\gamma$ in (c), respectively. (d) Normalized second-order correlation of the transition $|e\rangle\rightarrow|a\rangle$ as a function of delay $\tau$. The red solid line, blue dashed line, and green dot line correspond to the cases in (a)-(c), respectively. As a contrast, normalized second-order correlation of two-level system in the weak excitation regime is shown by black open circles. 
\label{Fig-2}}
\end{figure}
The intensity of fluorescence from the transition $|e\rangle\rightarrow|a\rangle$ in the steady-state limit is determined by the steady-state population of the excited state $\langle\sigma_{ee}\rangle = 2\Omega^2\Omega_r^2/(\Omega^4 + 2\Omega_r^2(2\gamma^2 + \Omega^2 + 2\Omega_r^2))$ (see Appendix~\ref{App-2}). The corresponding fluorescence spectrum is described by $S_{ea}(\omega_s)=\gamma_2 Re\int_0^{\infty}d\tau \lim\limits_{t\rightarrow\infty}\langle \sigma_{ea}(t)\sigma_{ae}(t+\tau)\rangle e^{i\omega_s \tau}$, where $\omega_s=\omega-\omega_L$ denotes the fluorescence frequency in the rotating frame at the driving frequency $\omega_L$.
One can see from Figs.~\ref{Fig-2}(a)-(c) that different spectral structures are exhibited for different Rabi frequency $\Omega_r$, which can be understood through the effective two-level system shown in Fig.~\ref{Fig-1}(b).
As mentioned above, the effective decay $|g\rangle\rightarrow|a\rangle$ corresponds to the two-photon process composed of the transition $|g\rangle\rightarrow|e\rangle$ driven by laser and the spontaneous decay $|e\rangle\rightarrow|a\rangle$.
It can be seen that the parameters adopted in Figs.~\ref{Fig-2}(a) and (c) correspond to the effective weak excitation regime, i.e., $\Omega_r\ll  \gamma^{*}$. Therefore, the fluorescence spectrum of the effective decay $|g\rangle\rightarrow|a\rangle$ exhibits a single-peak structure, whose spectral width is determined by the effective linewidth $\gamma^{*}$.
And because the laser frequency is constant, the transition $|e\rangle\rightarrow|a\rangle$ exhibits the same spectral structure as that in the effective decay $|g\rangle\rightarrow|a\rangle$ (see Appendix~\ref{App-3} for details).

Similarly, the parameters adopted in Fig.~\ref{Fig-2}(b) correspond to the effective strong excitation regime, i.e., $\Omega_r\gg \gamma^{*}$. Therefore, the fluorescence spectrum of the effective decay $|g\rangle\rightarrow|a\rangle$ exhibits a Mollow-like triplet~\cite{Mollow1969_Power, Wu1975_Investigation}, where the spectral widths of the three peaks depend on  $\gamma^{*}$ and the two side peaks are located at $\omega_L \pm 2\Omega_r$ (the proof is in Appendix~\ref{App-3}).
Accordingly, the transition $|e\rangle\rightarrow|a\rangle$ in the original three-level system exhibits the same Mollow-like triplet (see Appendix~\ref{App-3} for details).
Anyhow, all of the fluorescent components are almost focused on a bandwidth determined by the effective linewidth $\gamma^{*}$ and the sidebands in $\omega_L \pm 2\Omega_r$. This bandwidth is smaller than the natural linewidth of the emitter and can even be narrowed arbitrarily by reducing the intensities of the two external fields as shown in Fig.~\ref{Fig-2}(c).

\subsection{Statistical Property of Fluorescent Photons.}

The statistical property of the fluorescence emitted by the transition $|e\rangle\rightarrow|a\rangle$ can be reflected by the normalized second-order correlation $g^{(2)}_{ea}(\tau)=\lim\limits_{t\rightarrow\infty} \langle \sigma_{ea}(t)\sigma_{ea}(t+\tau)\sigma_{ae}(t+\tau)\sigma_{ae}(t)\rangle/\langle \sigma_{ee}(t)\rangle^2$ as shown in Fig.~\ref{Fig-2}(d). It reveals that the correlation $g^{(2)}_{ea}(\tau)$ remains a value very close to zero even when the delay $\tau$ is significantly larger than the lifetime of the emitter. In contrast, the normalized second-order correlation function of the two-level system increases rapidly with the increase of the delay. 
This feature of the correlation function can be understood according to the transition loop indicated in Fig.~\ref{Fig-1}(a). The emitter is driven from the ground state $|g\rangle$ to the excited state $|e\rangle$, and then enters into the ground state $|a\rangle$ after emitting the first photon from the spontaneous decay $|e\rangle\rightarrow|a\rangle$. The above transitions can be equivalent to the decay in the effective two-level system, and the corresponding decay rate is determined by $\gamma^{*}$.
After a long time determined by $\Omega_r$, the emitter is driven from state $|a\rangle$ to state $|g\rangle$.
Subsequently, the emitter can be excited again by the laser and then emit the second target photon.
Consequently, it is understandable that the correlation function $g^{(2)}_{ea}(\tau)$ remains zero for a long delay whose duration is determined by the effective decay rate $\gamma^{*}$ and the Rabi frequency $\Omega_r$ (the proof is in Appendix~\ref{App-4}).

\subsection{Detection Response.}

\begin{figure}
\centering\includegraphics[draft=false, width=1\columnwidth]{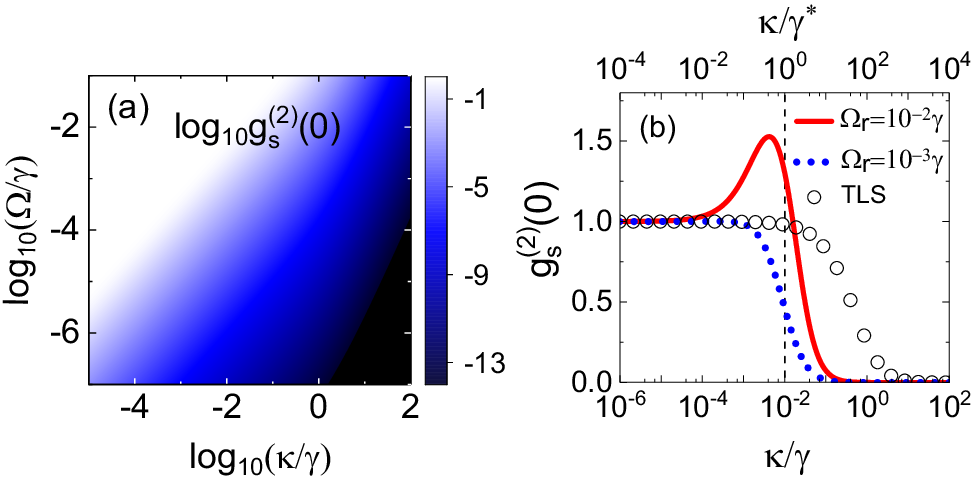}
\caption{Normalized two-photon correlations of the detector for the transition $|e\rangle\rightarrow|a\rangle$ (a) as a function of $\kappa$ and $\Omega$ with $\Omega_r=\Omega$, and (b) as a function of $\kappa$ ($\kappa/\gamma^*$), with $\Omega=10^{-1}\gamma $ and thus $\gamma^*=10^{-2}\gamma$. Normalized two-photon correlation of the detector for the two-level system in the weak excitation regime is shown as a contrast (black open circles).  
\label{Fig-3}}
\end{figure}
Single-photon character is the global property of fluorescence including all spectral components.
According to Refs.~\cite{Hanschke2020_Origin, Phillips2020_Photon}, only when the bandwidth of detector (or filter) is much larger than the linewidth of fluorescent single photons so that the dominant fluorescent components are included, can single-photon character be maintained when detected.  
Therefore, to verify the realization of the subnatural linewidth single photons, we consider the response of the fluorescence from transition $|e\rangle\rightarrow |a\rangle$ on the detector with a certain bandwidth, which can be described by the zero-delay two-photon correlation~\cite{Valle2012_Theory}
\begin{equation}
g_s^{(2)}(0) =\lim\limits_{t\rightarrow\infty}  \frac{\langle s^{\dag}(t)s^{\dag}(t)s(t)s(t) \rangle}{\langle s^{\dag}(t)s(t) \rangle ^2}.
\label{Eq-4}
\end{equation}

In Fig.~\ref{Fig-3}, we show the correlation $g_s^{(2)}(0)$ as a function of the detection bandwidth and the intensities of the external fields.
It reveals that when the detection bandwidth approaches the natural linewidth of the emitter, an excellent single-photon response for the fluorescence can appear on the detector, i.e., $g_s^{(2)}(0) \rightarrow 0$.
In contrast, for the fluorescence from a normal two-level system in the weak excitation regime where the main fluorescent components are focused on an ultranarrow coherent peak, the two-photon correlation on the detector deviates significantly from zero at the same bandwidth as shown in Fig.~\ref{Fig-3}(b), implying the destruction of single-photon character of fluorescence~\cite{Hanschke2020_Origin, Phillips2020_Photon, Carreno2018_Joint}.
According to the earlier discussion on the spectral and statistical properties of the fluorescence from transition $|e\rangle\rightarrow |a\rangle$, the physical origin of the excellent single-photon response on the detector can be understood.
In the weak excitation regime, almost all the spectral components of fluorescence are concentrated on a bandwidth that is significantly smaller than the natural linewidth of the emitter.
Therefore, when the detection bandwidth approaches the natural linewidth of the emitter, all spectral components of the fluorescence are proportionally responded to by the detector. Accordingly, the perfect single-photon character of the fluorescence is well preserved.
In turn, the excellent single-photon response for the quantum source on the detector also confirms the previous conclusion that the dominant components of fluorescence are concentrated on a bandwidth significantly smaller than the natural linewidth of the emitter.

Besides, one can see in Fig.~\ref{Fig-3}(a) that even if the detector bandwidth is much smaller than the natural linewidth of the emitter, i.e., $\kappa\ll\gamma$, the excellent single-photon response on the detector can still be maintained by manipulating the external fields.
Therefore, it can be concluded that the subnatural-linewidth single photons are achieved.
Moreover, the linewidth of the single photons can be manipulated easily by adjusting the intensities of external coherent fields, which dramatically facilitates the experimental implementation of this single-photon scheme, and has potential applications in extensive quantum technology fields.

The detection response can also be understood based on the effective level system. 
For a normal two-level system, when detection bandwidth is smaller than or approaches the natural linewidth, the destructive interference between coherent and incoherent components of the fluorescence is destroyed because a fraction of the incoherent component is removed by  filtering~\cite{Hanschke2020_Origin, Phillips2020_Photon, Carreno2018_Joint}. Consequently, the single-photon character is spoiled. 
Similarly, when the detection bandwidth $\kappa$ is smaller than or approaches the effective linewidth $\gamma^*$ of the effective level system shown in Fig.~\ref{Fig-1}(b), the single-photon character of the fluorescence is spoiled as shown in Fig.~\ref{Fig-3}(b).   
We can infer that only when $\kappa\gg \gamma^* + 4\Omega_r$ according to the earlier discussion, can the single-photon character be maintained.  

In addition, according to the effective level system shown in Fig.~\ref{Fig-1}(b), we see that when $\Omega_r\sim\gamma^{*}$  or $\Omega_r\gg\gamma^{*}$, the re-excitation rate of the state $|g\rangle$ approaches or exceeds the decay rate $\gamma^{*}$ at which the photon is emitted by the effective decay $|g\rangle\rightarrow |a\rangle$. Therefore, the emitted photons, which are actually the photons from the decay $|e\rangle\rightarrow |a\rangle$, exhibit a clustered temporal distribution, i.e., $g^{(2)}_{ea}(\tau)>1$, as shown by Fig.~\ref{fig-3LS_g2tNS}(b) in Appendix~\ref{App-4}. Since the second-order correlation $g^{(2)}_{ea}(\tau)$ of the emitter’s transition is linked to the two-photon correlation $g^{(2)}_{s}(0)$ on the detector, this temporal effect of photons is detected for a proper detection bandwidth, resulting in a bunching effect in the detector, i.e., $g_s^{(2)}(0)>0$, as shown by the red solid line in Fig.~\ref{Fig-3}(b) (A detailed explanation is provided in Appendix~\ref{App-4}).

\section{General condition for generating subnatural-linewidth single photons.}
\label{Sec-3}

According to Fig.~\ref{Fig-1}(a), we can see that there exists a long-period transition loop composed of the transitions $|g\rangle\rightarrow|e\rangle \xrightarrow{\rm{emission}} |a\rangle$ and $|a\rangle\rightarrow|g\rangle$, which dominates the successive emissions from the transition $|e\rangle\rightarrow|a\rangle$.
Specifically, the emitter in the excited state $|e\rangle$ emits the first photon from the transition $|e\rangle\rightarrow|a\rangle$ and enters into the state $|a\rangle$. 
Next, the emitter is driven to the state $|g\rangle$ by the coherent field F2, and then is driven to the excited state $|e\rangle$ by the laser field F1. Subsequently, the second photon from the transition $|e\rangle\rightarrow|a\rangle$ is emitted.
Noteworthily, this transition loop totally dominates the successive emissions from the transition $|e\rangle\rightarrow|a\rangle$, because the probabilities of the other transition loops are negligible.
And due to $\Omega_r\ll\gamma $, the state $|a\rangle$ is the metastable state or the so-called shelving state~\cite{Nagourney1986_Shelved, Sauter1986_Observation, Bergquist1986_Observation, Plenio1998_The-quantum-jump}, which can remain for a time much longer than the lifetime of the excited state.  Therefore, the time interval between two successive photons from the transition $|e\rangle\rightarrow|a\rangle$ is sufficiently long compared to the lifetime of the emitter.

According to the Heisenberg uncertainty principle, when the detection bandwidth $\kappa$ is smaller than the natural linewidth of the emitter, the indeterminacy in the time resolution of detection $1/\kappa$ is larger than the natural lifetime of the emitter.
And within a large delay determined by indeterminacy $1/\kappa$, the normalized probability of reemission from the transition $|e\rangle\rightarrow|a\rangle$ can remain very close to zero as shown in Fig.~\ref{Fig-2}(c), so an excellent single-photon response arises on the detector with a narrow bandwidth.

It is the direct indication of the subnatural-linewidth single photons to induce an excellent single-photon response on the detector with a bandwidth approaching or smaller than the natural linewidth of emitter. In terms of time, the detection bandwidth $\kappa$ corresponds to the indeterminacy $1/\kappa$ in time, then the decrease of single-photon linewidth means the increase of the average time interval between single photons. In fact, in the spectrum of fluorescence, i.e., fluorescent single photons, the contributions of multiple transition loops may be included. Among them, the transition loops without a metastable state would contribute to broad spectral components, leading to the broad linewidth of single photons. 
Therefore, we can conclude that the general condition for generating subnatural-linewidth single photons is that the successive emission process of the target photons is totally dominated by the transition loop with a metastable state. The validity of this condition is comprehensible by means of detection response and can be verified by various emitter systems shown in Appendix~\ref{App-5} and Appendix~\ref{App-6}.

\section{Single-photon emission rate.}
\label{Sec-4}

\subsection{Limit of single-photon emission rate.} 
\label{Sec-4a}

The linewidth of a single-photon wavepacket corresponds to its frequency distribution range $\Delta\omega$. According to the uncertainty relation $\Delta\omega \Delta t \geq 1$, the time distribution range of the single-photon wavepacket satisfies $ \Delta t \geq 1/\Delta\omega $. 
This means that if all frequency components of a single photon are concentrated within a width $\Delta\omega$, at most one photon can exist within the time range $1/\Delta\omega$. Therefore, for a field satisfying single-photon statistics, the number of photons per unit time must satisfy
\begin{equation}
	I_{\rm{s}} \equiv \dfrac{1}{\Delta t} \leq \Delta\omega .    
\label{Eq-5}   
\end{equation}
According to Eq.~(\ref{Eq-5}), the single-photon linewidth determines the upper limit of the single-photon emission rate. Therefore, when the single-photon linewidth is small, the single-photon emission rate of the emitter would inevitably be small, regardless of the type of emitter. 

\subsection{Emission rate of subnatural-linewidth single photons.} 

\begin{figure}[htbp]
	\centering\includegraphics[draft=false, width=0.6\columnwidth]{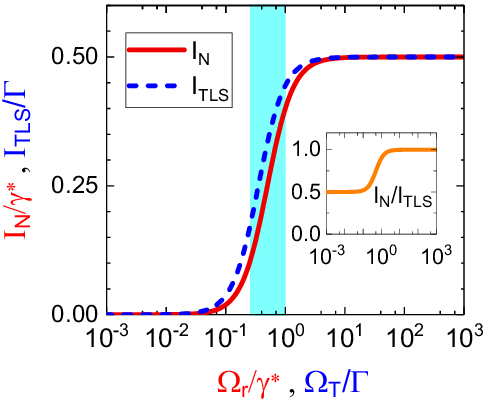}
	\caption{ Emission rates of the subnatural-linewidth single photons  (red solid line) and two-level system (blue dashed line) as functions of Rabi frequencies $\Omega_r$ and $\Omega_T$, respectively. In the region with the blue background, the emission rate of the subnatural-linewidth single photons almost reaches the upper limit. The inset shows the ratio of these two single-photon emission rates, providing $\gamma^*=\Gamma$.\label{fig-inten}}
\end{figure}

To analyze the emission rate of the subnatural-linewidth single photons, we first consider the normal two-level system for comparison. Two-level system is the most general fluorescence emitter, whose single-photon emission rate can be expressed as
\begin{equation}
	I_{\rm{TLS}}=\Gamma \frac{4\Omega_T^2}{\Gamma^2 + 8\Omega_T^2},    
\label{Eq-6}
\end{equation}
where $\Gamma$ and $\Omega_T$ denote the natural linewidth of the two-level system and the Rabi frequency of the laser field, respectively. 
In the weak excitation regime, i.e., $\Omega_T\ll \Gamma$, the single-photon emission rate is 
\begin{equation}
	I_{\rm{TLS}}=\frac{4\Omega_T^2}{\Gamma}\ll \Gamma.     
\label{Eq-7}
\end{equation}
With the increase of $\Omega_T$, the single-photon emission rate increases and gradually approaches $\Gamma$. 
In the strong excitation regime, i.e., $\Omega_T\gg \Gamma$, the single-photon emission rate reaches the maximum value   
\begin{equation}
I_{\rm{TLS}}=\frac{\Gamma}{2},     
\label{Eq-8}
\end{equation}
which is determined by the natural linewidth $\Gamma$.

Moreover, we know that when $\Omega_T \sim \Gamma$, the sidebands in the fluorescence are not resolved, and thus, the bandwidth of all fluorescent components, i.e., the linewidth of fluorescent single photons, is approximately equal to the natural linewidth $\Gamma$. As shown by the blue dashed line in Fig.~\ref{fig-inten}, the single-photon emission rate also approaches $\Gamma$ when $\Omega_T\sim\Gamma$. Therefore, we can conclude that for the two-level emitter, the single-photon emission rate can almost reach the upper limit determined by the single-photon linewidth according to Eq.~(\ref{Eq-5}). 

For the emitter shown in Fig.~\ref{Fig-1}(a), the emission rate $I_{N}=\gamma_2 \langle\sigma_{ee}\rangle$ of the subnatural-linewidth single photons can be obtained as (see Appendix~\ref{App-2})
\begin{equation}
	I_{N}=\gamma \frac{2  \Omega ^2 \Omega _r^2}{\Omega^4 + 2\Omega_r^2(2\gamma^2 + \Omega^2 + 2\Omega_r^2)}.    \label{Eq-9}
\end{equation}
For the weak excitation regime we focus on, i.e., $\Omega, \Omega_r\ll\gamma$, one can understand the emission rate of the subnatural-linewidth single photons by means of the effective two-level system similar to that in the previous sections. Specifically, Eq.~(\ref{Eq-9}) can be reduced to a function of the effective linewidth $\gamma^*$ and the Rabi frequency $\Omega_r$
\begin{equation}
	I_{N}=\gamma^* \frac{2  \Omega _r^2}{  \gamma ^{*2} + 4 \Omega _r^2}.     
\label{Eq-10}
\end{equation}
Further, in the effective weak excitation regime, i.e., $\Omega_r\ll \gamma^*$ , the single-photon emission rate is 
\begin{equation}
	I_{N}= \frac{2\Omega_r^2}{\gamma^*}\ll\gamma^*.     
\label{Eq-11}
\end{equation}
In the effective strong excitation regime, i.e., $\Omega_r\gg \gamma^*$, the single-photon emission rate reaches the maximum value 
\begin{equation}
I_{N}= \frac{\gamma^*}{2},    
\label{Eq-12}
\end{equation}
which is determined by the effective linewidth $\gamma^*$.

One can deduce from Eqs.~(\ref{Eq-10})-(\ref{Eq-12}) and Fig.~\ref{fig-inten} that the rules followed by the emission rate of the subnatural-linewidth single photons are very similar to that in the two-level system, except that the natural linewidth $\Gamma$ and the Rabi frequency $\Omega_T$ are replaced by the effective linewidth $\gamma^*$ and the Rabi frequency $\Omega_r$. Measured in terms of actual single-photon linewidth, i.e., providing $\gamma^*=\Gamma$, the emitter of subnatural-linewidth single photons we propose has almost the same emission rate as the most general two-level emitter. 
For a visual demonstration, we show the ratio $ I_{N} /I_{TLS}$ as functions of Rabi frequency in the inset of Fig.~\ref{fig-inten}. It reveals that in the (effective) weak excitation regime, the emission rate of the subnatural-linewidth single photons can reach half of that of the two-level emitter. As the Rabi frequency increases, the emission rate of the subnatural-linewidth single photons gradually reaches the maximum value as that in two-level emitter, as demonstrated in Eqs.~(\ref{Eq-8}) and ~(\ref{Eq-12}). 

Significantly, similar to the case in two-level emitter, the sidebands of the fluorescence are not resolved when $\Omega_r \sim \gamma^*$, and thus, the linewidth of fluorescent single photons is largely determined by the effective linewidth $\gamma^*$. Besides, the single-photon emission rate also approaches $\gamma^*$ when $\Omega_r \sim \gamma^*$ as shown by the red solid line in the region with blue background in Fig.~\ref{fig-inten}. Therefore, we can conclude that for the subnatural-linewidth single photons we propose, the emission rate can almost reach the upper limit determined by its single-photon linewidth $\gamma^*$ according to Eq.~(\ref{Eq-5}). Moreover, since $\gamma^*=|\Omega|^2/\gamma$ is determined by the intensity of the external field, the performance of single photons can be conveniently manipulated while maintaining the maximum emission rate. 

The constraint between single-photon linewidth and emission rate can be understood in terms of the uncertainty principle followed by single-photon wave packets, as described in Sec.~\ref {Sec-4a}. Besides, this constraint also exhibits on the detection process. The narrow bandwidth $\kappa$ of the detector means a long response time $1/\kappa$. And for multiple incident photons occurring within this response time, the detector cannot distinguish the order of these photons, but consider that they are simultaneously incidents. Therefore, to ensure the single-photon response on the detector with a narrow bandwidth, the average time interval of successive single photons should be much longer than the response time of the detector, which limits the number of photons per unit time.

\section{Physical realization and discussion.}
\label{Sec-5}
Various emitters with $\Lambda$ or $\Lambda$-like structures can be used to generate single photons with subnatural linewidth.
For instance, as a system extensively investigated and employed in quantum technologies ~\cite{Volz2006_Observation, Hofmann2012_Heralded, Leent2020_Long-Distance}, the single ${}^{87}$Rb atom loaded into an optical dipole trap is suitable for the implementation of our proposed approach. 
Specifically, the closed hyperfine transition $F_{g}=1\rightarrow F_{e}=0$ of the $D_{2}$ line of ${}^{87}$Rb is driven by a circularly polarized light resonantly, and the transitions between the ground-state Zeeman sublevels are driven by the magnetic field perpendicular to the direction of light propagation simultaneously. Therefore, single photons with subnatural linewidth can be emitted (see Appendix~\ref{App-5} for more details).  
In addition, our approach can be implemented based on the $\Lambda$-level structure contained in the charged GaAs and InGaAs quantum dots in Voigt magnetic field~\cite{Dutt2005_Stimulated, He2015_Dynamically}, where a transition between an electron-spin ground state ($|\uparrow\rangle$ or $|\downarrow\rangle$) and trion excited state  ($|\downarrow\uparrow\Uparrow\rangle$ or $|\downarrow\uparrow\Downarrow\rangle$) can be driven by a laser resonantly, and the coherent transition between electron-spin ground states can be realized by a magnetic field or a stimulated Raman process. 
Both the frequencies and polarizations of the target single photons and the driving field can be different in our approach. Therefore, based on frequency and polarization filtering, as well as existing experimental techniques~\cite{Wang2019_Towards}, the laser background can be excluded efficiently to improve the quality of single photons.

Noteworthily, reduction of the linewidth of subnatural-linewidth single photons are relative to the natural linewidth of a given emitter. According to the general condition for generating subnatural-linewidth single photons, our study can be readily implemented based on various physical systems, e.g., atom, quantum dot, ion, molecule, and nitrogen-vacancy center, whose natural linewidths vary over a wide range from MHz to THz~\cite{Lounis2005_Single-photon}. Therefore, the specific linewidth and emission rate of the subnatural-linewidth single photons, which depend on the the specific kind of emitter, also vary over wide ranges. 
For instance, with the parameters shown in Fig.~\ref{Fig-2}(a), subnatural-linewidth single photons with a bandwidth of 1/10 natural linewidth are generated. In this case, if the single-photon scheme is implemented based on a quantum dot with a typical natural linewidth of the order of GHz, the linewidth and thus emission rate of the subnatural-linewidth single photon will be about 100 MHz, which are significantly larger than the linewidth and emission rate of the single photons emitted by an atom with a typical natural linewidth of the order of MHz. 
Therefore, we can conclude that ``subnatural linewidth'' does not denote ``narrow linewidth'', but rather breaks the limit imposed by natural linewidth on the linewidth of fluorescent single photons. In practice, subnatural-linewidth single photons may have a large linewidth and emission rate.

\section{Summary.}
\label{Sec-6}

We propose an approach for focusing all the spectral components of fluorescence on subnatural linewidth, and thus the single photons with subnatural linewidth can be emitted under the premise of inheriting the perfect single-photon character of fluorescence. Thereby, the limit imposed by the natural linewidth of the emitter's transition on the linewidth of fluorescent single photons is broken. Comparing to the constant natural linewidth of emitter, the single-photon linewidth can be easily manipulated over a broad range by varying the intensities of external fields. 
Further, we reveal that the general condition for generating single photons with subnatural linewidth utilizing the atom or atom-like system is that the successive emission process of the target photons is totally dominated by the transition loop with a metastable state.  
Benefiting from this revelation, this proposed approach can be generalized in various physical systems with $\Lambda$-shape and similar energy structures. 
Since the distribution of single-photon wave packets in time and frequency has to satisfy the uncertainty relation, the upper limit of the emission rate imposed by the single-photon linewidth is inevitable, both for the conventional single photons and for the subnatural-linewidth single photons. However, we demonstrate that the emission rate of the subnatural-linewidth single photons can reach its theoretical upper limit. 

In addition, the single photons with narrow or even ultranarrow linewidth, which have long coherence time, can also be obtained by reducing the intensities of external fields. 
For the excellent tunability of the single-photon linewidth and emission rate, it is possible to find the best trade-off between these parameters according to the requirement in a specific situation.
Consequently, we see that the capability to manipulate the linewidth and emission rate of single photons in various atomic and atom-like systems can be significantly enhanced based on our research.  This research will significantly facilitate the insight into the quantum nature of single photons and resonance fluorescence, and have extensive applications in quantum optics, photonics, and quantum technologies due to the outstanding performance of subnatural and narrow linewidth single photons.

\appendix

\section{Effective motion equation}
\label{App-1}

In the limit of the vanishing coupling between the emitter and detector, the detector is included as a passive object in the motion equation~(\ref{Eq-2}), and its backaction on the emitter is neglected. Therefore, we can separately solve and discuss the motion equation describing the evolution of the emitter, that is, 
\begin{equation}
	\dot{\rho_{\sigma}}=-i [\textit{H}_{A},\rho_{\sigma}]  + \gamma_{1} \mathcal{D} [\sigma_{ge}]\rho_{\sigma} + \gamma_{2} \mathcal{D}[\sigma_{ae}]\rho_{\sigma},     
	\label{Eq-a1}
\end{equation}
where $\textit{H}_{A}= \Omega \sigma_{eg} + \Omega_r \sigma_{ga}  + \rm{H.c} $ is the Hamiltonian of the emitter, and $\rho_{\sigma}$ represents the reduced density matrix of the emitter. Moreover, in the weak excitation regime of the emitter, i.e., $\Omega, \Omega_r \ll \gamma_1, \gamma_2$, the excited state can be adiabatically eliminated using the second-order perturbation theory. Specifically, the motion equations associated with the excited state is  
\begin{align}
	\dot{\rho}_{ee} &=-(\frac{\gamma_1}{2}+\frac{\gamma_2}{2})\rho_{ee} -i\Omega\rho_{ge} + i\rho_{eg}\Omega^*  , \nonumber  \\
	\dot{\rho}_{eg} &= -(\frac{\gamma_1}{2}+\frac{\gamma_2}{2})\rho_{eg} + i\Omega\rho_{ee} - i\Omega\rho_{gg} + i\rho_{ea}\Omega_r ,  \nonumber  \\
	\dot{\rho}_{ea} &= -(\frac{\gamma_1}{2}+\frac{\gamma_2}{2})\rho_{ea} - i\Omega\rho_{ga} + i\rho_{eg}\Omega_r^* ,
	\label{Eq-a2}
\end{align}
where $\rho_{mn}= \langle \sigma_{nm} \rangle $ with $m,n = g,a,e$ . Setting the left side of the equal sign of Eq.~(\ref{Eq-a2}) to zero in order to discard the rapid evolution effect, we can therefore deduce that
\begin{align}
	\rho_{ge}(t) &\approx \frac{2i\Omega^*}{\gamma_1+\gamma_2} \rho_{gg}(t)  ,  \nonumber  \\
	\rho_{ae}(t) &\approx \frac{2i\Omega^*}{\gamma_1+\gamma_2} \rho_{ag}(t) ,    \nonumber  \\
	\rho_{ee}(t) &\approx \frac{4|\Omega|^2}{(\gamma_1+\gamma_2)^2} \rho_{gg}(t)  .
	\label{Eq-a3}
\end{align}

The motion equations of the ground states are 
\begin{align}
	\dot{\rho}_{gg} &= \gamma_1\rho_{ee} + i\Omega\rho_{ge} + i\Omega_r\rho_{ga} - i\Omega^* \rho_{eg} - i\Omega_r^* \rho_{ag}  , \nonumber  \\
	\dot{\rho}_{aa} &= \gamma_2\rho_{ee} - i\Omega_r\rho_{ga} + i\Omega_r^* \rho_{ag}  , \nonumber  \\
	\dot{\rho}_{ga} &= - i\Omega^* \rho_{ea} - i\Omega_r^* \rho_{aa} + i\Omega_r^* \rho_{gg} . 
	\label{Eq-a4}
\end{align}
Substituting Eqs.~(\ref{Eq-a3}) into (\ref{Eq-a4}), we obtain the motion equations containing only the degrees of freedom of the ground states, which can be equivalently written in Lindblad form as
\begin{equation}
	\dot{\rho}_g = -i[H_{T}, \rho_g] + \mathcal{L}_{\textrm{eff}}\rho_g  ,
	\label{Eq-a5}
\end{equation}
with Hamiltonian $\textit{H}_{T}= \Omega_r \sigma_{ga}  + \rm{H.c} $ and the terms $\mathcal{L}_{\textrm{eff}}\rho_g$ denoting effective decays obtained from the adiabatic elimination of the excited state 
$\mathcal{L}_{\textrm{eff}}\rho_g = \gamma_1^* \mathcal{D}[\sigma_{gg}]\rho_g + \gamma_2^* \mathcal{D}[\sigma_{ag}]\rho_g $. Effective linewidths satisfy $\gamma^{*}_{i}=4\gamma_i|\Omega|^2/(\gamma_1+\gamma_2)^2$ with $\textit{i}=1,2$.

\section{Steady-state solution of emitter}
\label{App-2}

By solving Eq.~(\ref{Eq-a1}) under steady-state conditions, we get the analytical solutions of the density matrix elements of the emitter
\begin{align}
	\tilde\rho_{gg} &= \frac{\Omega_r^2(2\gamma^2 + \Omega^2 + 2\Omega_r^2)}{M} ,    
    \nonumber    \\
 \tilde\rho_{aa} &= \frac{\Omega^4 + \Omega_r^2(2\gamma^2 - \Omega^2 + 2\Omega_r^2)}{M} ,     \nonumber    \\
	\tilde\rho_{ee} &= \frac{2\Omega^2\Omega_r^2}{M} ,       \qquad	  
	\tilde\rho_{ge} = \frac{2i\gamma\Omega\Omega_r^2}{M} ,     \nonumber    \\
	\tilde\rho_{ae} &= \frac{-\Omega^3\Omega_r + 2\Omega\Omega_r^3}{M} ,   \qquad	 
	\tilde\rho_{ga} = \frac{-i\gamma\Omega^2\Omega_r}{M} ,
	\label{Eq-a6}
\end{align}
where $M=\Omega^4 + 2\Omega_r^2(2\gamma^2 + \Omega^2 + 2\Omega_r^2)$ (we assume $\gamma_1=\gamma_2=\gamma$ for simplicity). It can be seen that when the coherent field F2 is removed, the state $|a\rangle$ becomes a dark state on which all the populations are focused.

\section{Detailed solution to the single-photon spectrum}
\label{App-3}

\begin{figure*}[htbp]
\centering\includegraphics[draft=false, width=1.35\columnwidth]{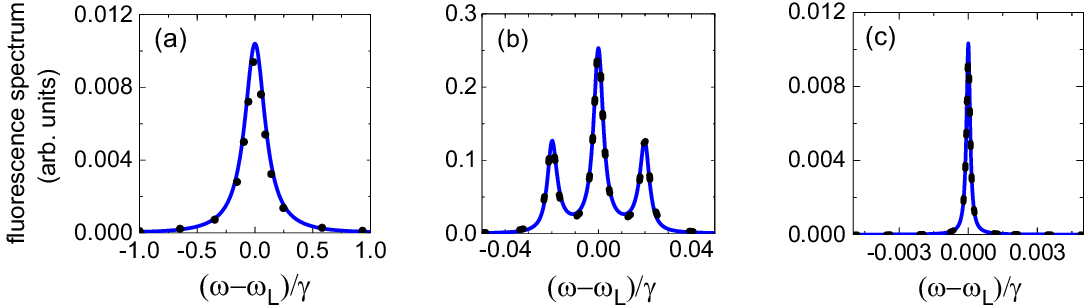}
\caption{Comparison of analytical and numerical results of fluorescence spectrum. The analysis results (solid line) of the fluorescence spectrum (incoherent component) according to Eq.~(\ref{Eq-a10}) in (a) and (c) and Eq.~(\ref{Eq-a16}) in (b) are compared with the numerical results (dotted line) to verify the accuracy of the theory in this section. The parameters applied in (a)-(c) are the same as that in Figs. ~\ref{Fig-2} (a)-(c), respectively.
\label{fig-3LS_spectrumNS}}
\end{figure*}

The steady-state fluorescence spectrum for the transition $|e\rangle\rightarrow|a\rangle$ in the far-field zone is given by
\begin{equation}
	S_{ea}(\omega_s)=\gamma_2 Re\int_0^{\infty}d\tau \lim\limits_{t\rightarrow\infty}\langle \sigma_{ea}(t)\sigma_{ae}(t+\tau)\rangle e^{i\omega_s \tau}  ,
	\label{Eq-a7}
\end{equation}
with $\omega_s = \omega - {\omega _L}$. 
According to the quantum regression theorem~\cite{Lax1963_Formal, Carmichael1993_An}, the evolution of $ \left\langle {{\sigma _{ea}}(t){\sigma _{mn}}(t + \tau )} \right\rangle$ with $\tau$ and the evolution of the density matrix element $\rho_{nm}(t)$ of the emitter with $t$ obey the same equation. Therefore, based on the master equation~(\ref{Eq-a1}), we can obtain the following equations with the definition $Y_{mn}(\tau) \equiv \left\langle {{\sigma _{ea}}(t){\sigma _{mn}}(t + \tau )} \right\rangle$ 
\begin{align}
	{\partial _\tau }\!{Y_{gg}}\!\!(\!\tau\!) &= {\gamma _1}{Y_{ee}}\!(\!\tau\!) + i\Omega {Y_{eg}}\!(\!\tau\!) - i{\Omega ^*}{Y_{ge}}\!(\!\tau\!) + i{\Omega _r}{Y_{ag}}\!(\!\tau\!) - i\Omega _r^*{Y_{ga}}\!(\!\tau\!),  \nonumber  \\
	{\partial _\tau }\!{Y_{ag}}\!(\!\tau\!) &=  - i\Omega _r^*{Y_{aa}}\!(\!\tau\!) - i{\Omega ^*}{Y_{ae}}\!(\!\tau\!) + i\Omega _r^*{Y_{gg}}\!(\!\tau\!),  \nonumber  \\
	{\partial _\tau }\!{Y_{eg}}\!(\!\tau\!) &=  - \left( {\frac{{{\gamma _1}}}{2} + \frac{{{\gamma _2}}}{2}} \right){Y_{eg}}\!(\!\tau\!) - i{\Omega ^*}{Y_{ee}}\!(\!\tau\!) + i{\Omega ^*}{Y_{gg}}\!(\!\tau\!) - i\Omega _r^*{Y_{ea}}\!(\!\tau\!),  \nonumber  \\
	{\partial _\tau }\!{Y_{ga}}\!(\!\tau\!) &= i\Omega {Y_{ea}}\!(\!\tau\!) + i{\Omega _r}{Y_{aa}}\!(\!\tau\!) - i{\Omega _r}{Y_{gg}}\!(\!\tau\!),  \nonumber  \\
	{\partial _\tau }\!{Y_{aa}}\!(\!\tau\!) &= {\gamma _2}{Y_{ee}}\!(\!\tau\!) - i{\Omega _r}{Y_{ag}}\!(\!\tau\!) + i\Omega _r^*{Y_{ga}}\!(\!\tau\!),  \nonumber  \\
	{\partial _\tau }\!{Y_{ea}}\!(\!\tau\!) &=  - \left( {\frac{{{\gamma _1}}}{2} + \frac{{{\gamma _2}}}{2}} \right){Y_{ea}}\!(\!\tau\!) + i{\Omega ^*}{Y_{ga}}\!(\!\tau\!) - i{\Omega _r}{Y_{eg}}\!(\!\tau\!),  \nonumber  \\
	{\partial _\tau }\!{Y_{ge}}\!(\!\tau\!) &=  - \left( {\frac{{{\gamma _1}}}{2} + \frac{{{\gamma _2}}}{2}} \right){Y_{ge}}\!(\!\tau\!) + i\Omega {Y_{ee}}\!(\!\tau\!) - i\Omega {Y_{gg}}\!(\!\tau\!) + i{\Omega _r}{Y_{ae}}\!(\!\tau\!),  \nonumber  \\
	{\partial _\tau }\!{Y_{ae}}\!(\!\tau\!) &=  - \left( {\frac{{{\gamma _1}}}{2} + \frac{{{\gamma _2}}}{2}} \right){Y_{ae}}\!(\!\tau\!) - i\Omega {Y_{ag}}\!(\!\tau\!) + i\Omega _r^*{Y_{ge}}\!(\!\tau\!),  \nonumber  \\
	{\partial _\tau }\!{Y_{ee}}\!(\!\tau\!) &=  - \left( {{\gamma _1} + {\gamma _2}} \right){Y_{ee}}\!(\!\tau\!) - i\Omega {Y_{eg}}\!(\!\tau\!) + i{\Omega ^*}{Y_{ge}}\!(\!\tau\!).
	\label{Eq-a7-1}
\end{align}
Due to ${Y_{mn}}(0) = \langle {\sigma _{ea}}(t){\sigma _{mn}}(t)\rangle$, we obtain the nonzero initial values of $Y_{mn}(0)$ in the limit $\rm{t} \to \infty $ as
\begin{equation}
{Y_{ag}}(0) = {\tilde \rho _{ge}},   \qquad	 
{Y_{aa}}(0) = {\tilde \rho _{ae}},   \qquad	 
{Y_{ae}}(0) = {\tilde \rho _{ee}},
\label{Eq-a7-2}
\end{equation}
where $\tilde \rho_{nm}$ is the steady-state density matrix element in Eq.~(\ref{Eq-a6}). 
We next introduce the Laplace transform of the correlation $Y_{mn}(\tau)$
\begin{equation}
	{Y_{mn}}\left( s \right) \equiv \int_0^{\infty}d\tau  {Y_{mn}}\left( \tau  \right){e^{ - s\tau }} ,
\label{Eq-a7-3}
\end{equation}
with $s = -i\omega_s$. By performing the Laplace transform on Eq.~(\ref{Eq-a7-1}), we can obtain a set of linear equations about $Y_{mn}(s)$. Based on this set of equations and Eq.~(\ref{Eq-a7}), the solution of the steady-state fluorescence spectrum can be obtained.

Consistent with the discussion in the main text, we focus on the weak excitation regime, i.e., $\Omega,\Omega_r\ll\gamma_1,\gamma_2$.
In the effective weak excitation regime, i.e., $\Omega_r\ll\gamma^{*}$, the analytical expression of the fluorescence spectrum can be obtained by the perturbation method as
\begin{equation}
	S^{ew}_{ea}(\omega_s) = S^{ew}_{c}(\omega_s) + S^{ew}_{n}(\omega_s) + S^{ew}_{b}(\omega_s).
	\label{Eq-a8}
\end{equation}
Here $S^{ew}_{c}(\omega_s)$ represents the coherent component of the fluorescence and is given by
\begin{equation}
	S^{ew}_{c}(\omega_s) = \gamma |\tilde\rho_{ae}|^2\delta(\omega_s).
	\label{Eq-a9}
\end{equation}
$S^{ew}_{n}(\omega_s)$ represents the incoherent component with a narrow linewidth of the fluorescence and is given by
\begin{eqnarray}
	S^{ew}_{n}(\omega_s) = \frac{-i\tilde\rho_{ge}\Omega^2 + \tilde\rho_{ae}\gamma \Omega_r}{\Omega}\frac{\gamma^{*}}{\omega_s^2+\gamma^{*2}}.
	\label{Eq-a10}
\end{eqnarray}
$S^{ew}_{b}(\omega_s)$ represents the incoherent component with a broad linewidth of the fluorescence and is given by
\begin{eqnarray}
	S^{ew}_{b}(\omega_s) = -\tilde\rho_{ae}\Omega\Omega_r\frac{1}{\omega_s^2+\gamma^2}-\frac{2i\tilde\rho_{ge}\Omega^3}{\gamma}\frac{\gamma^2-\omega_s^2}{(\gamma^2+\omega_s^2)^2}.
	\label{Eq-a11}
\end{eqnarray}

It can be obtained that the ratio between the intensities of the incoherent components with broad and narrow linewidths is
\begin{eqnarray}
	\frac{\int_{\infty}^{\infty}d\omega_s S^{ew}_{b}}{\int_{\infty}^{\infty}d\omega_s S^{ew}_{n}}  = \frac{\Omega^2}{\gamma^2},
	\label{Eq-a12}
\end{eqnarray}
which is much less than 1. Therefore, the incoherent component with broad linewidth is ignorable, and the analytical expression of the fluorescence spectrum is reduced as
\begin{eqnarray}
	S^{ew}_{ea}(\omega_s) = S^{ew}_{c}(\omega_s) + S^{ew}_{n}(\omega_s),
	\label{Eq-a13}
\end{eqnarray}
which is in good agreement with the exact numerical result according to Figs.~\ref{fig-3LS_spectrumNS}(a) and (c). We see that the fluorescence spectrum exhibits a single-peak structure whose spectral width determined by $\gamma^{*}$ is significantly smaller than the natural linewidth of the emitter $\gamma$.

In the effective strong excitation regime, i.e., $\Omega_r\gg\gamma^{*}$, the analytical expression of the fluorescence spectrum can be obtained as
\begin{eqnarray}
	S^{es}_{ea}(\omega_s) = S^{es}_{c}(\omega_s) + S^{es}_{n}(\omega_s) + S^{es}_{b}(\omega_s).
	\label{Eq-a14}
\end{eqnarray}
Here $S^{es}_{c}(\omega_s)$ represents the coherent component of the fluorescence and is given by
\begin{eqnarray}
	S^{es}_{c}(\omega_s) = \gamma |\tilde\rho_{ae}|^2\delta(\omega_s).
	\label{Eq-a15}
\end{eqnarray}
$S^{es}_{n}(\omega_s)$ represents the incoherent component with a narrow linewidth of the fluorescence and is given by
\begin{equation}
	S^{es}_{n}\!(\!\omega_s\!) \!=\! -\!\frac{i}{4}\tilde\rho_{ge}\Omega (\frac{2\gamma^{*}}{\omega_s^2\!+\!\gamma^{*2}} \!+\! \frac{\gamma^{*}}{(\omega_s\!-\!2\Omega_r)^2+\gamma^{*2}} \!+\! \frac{\gamma^{*}}{(\omega_s\!+\!2\Omega_r)^2\!+\!\gamma^{*2}}).
	\label{Eq-a16}
\end{equation}
$S^{es}_{b}(\omega_s)$ represents the incoherent component with broad linewidth and is given by
\begin{eqnarray}
	S^{es}_{b}(\omega_s) = -\tilde\rho_{ae}\Omega\Omega_r\frac{1}{\omega_s^2+\gamma^2}.
	\label{Eq-a17}
\end{eqnarray}

It can also be obtained that the ratio between the intensities of the incoherent components with broad and narrow linewidths is
\begin{eqnarray}
	\frac{\int_{\infty}^{\infty}d\omega_s S^{es}_{b}}{\int_{\infty}^{\infty}d\omega_s S^{es}_{n}}  = \frac{\Omega^2-2\Omega_r^2}{2\gamma^2},
	\label{Eq-a18}
\end{eqnarray}
which is much less than 1. Therefore, the incoherent component with broad linewidth is ignorable, and the analytical expression of the fluorescence spectrum is reduced as
\begin{eqnarray}
	S^{es}_{ea}(\omega_s) = S^{es}_{c}(\omega_s) + S^{es}_{n}(\omega_s),
	\label{Eq-a19}
\end{eqnarray}
which is in good agreement with the exact numerical result according to Fig.~\ref{fig-3LS_spectrumNS}(b). We see that the fluorescence spectrum exhibits a Mollow-like triplet~\cite{Mollow1969_Power, Wu1975_Investigation}.
The spectral width of each peak also depends on the effective linewidth $\gamma^{*}$, thus is smaller than the natural linewidth $\gamma$ of the emitter.
Moreover, since the two sidebands arise at frequencies $\omega_L\pm 2\Omega_r$, one can infer that all the spectral components of fluorescence are concentrated in a bandwidth of $4\Omega_r$, which is smaller than the natural linewidth of the emitter as shown in Fig.~\ref{fig-3LS_spectrumNS}(b).

\section{Detailed solution and evolution of photonic statistical correlation}
\label{App-4}

\begin{figure}[htbp]
\centering\includegraphics[draft=false, width=1\columnwidth]{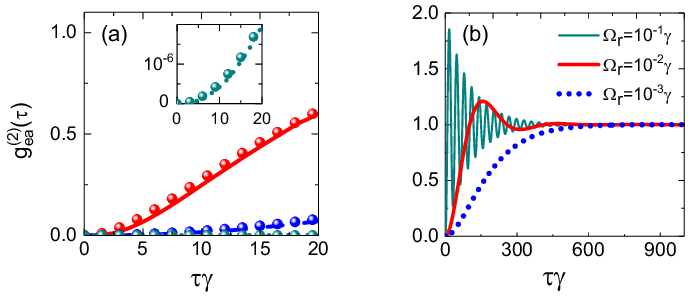}
\caption{(a) Comparison of analytical and numerical results of the correlation function $g^{(2)}_{ea}(\tau)$. The analytical results (solid circles) of the correlation function according to Eq.~(\ref{Eq-a20}) are compared with the numerical results shown in Fig.~\ref{Fig-2} (d) to verify the accuracy of the theory in this section. 
(b) Evolution of $g^{(2)}_{ea}(\tau)$ over a time scale greater than $\gamma^{*-1}$ with $\Omega=10^{-1}\gamma $ and thus $\gamma^*=10^{-2}\gamma$.
\label{fig-3LS_g2tNS}}
\end{figure}

In the weak excitation regime, the analytical expression of the normalized second-order correlation $g^{(2)}_{ea}(\tau)$ can be obtained according to the quantum regression theorem as
\begin{equation}
	g^{(2)}_{ea}(\tau)=1 - \cos(2\Omega_r\tau)e^{-\gamma^{*}\tau} - \frac{\Omega^2-2\Omega_r}{2\gamma\Omega_r}\sin(2\Omega_r\tau) e^{-\gamma^{*}\tau},
	\label{Eq-a20}
\end{equation}
which is in good agreement with the exact numerical result according to Fig.~\ref{fig-3LS_g2tNS}(b). We can see from Eq.~(\ref{Eq-a20}) that the time evolution of the second-order correlation $g^{(2)}_{ea}(\tau)$ depends on the effective linewidth $\gamma^{*}$ and Rabi frequency $\Omega_r$ of the coherent field F2, both of which are determined by the external coherent fields and thus can be easily manipulated experimentally.
In contrast, in the same parameter regime, the second-order correlation function of a normal two-level system (i.e., $g_{T}^{(2)}(\tau)=(1 - e^{-\frac{\Gamma}{2}\tau})^2$) depends entirely on the natural linewidth $\Gamma$, which cannot be modified arbitrarily.

In addition, we see from Fig.~\ref{fig-3LS_g2tNS}(b) that under the effective weak excitation regime, i.e., $\Omega_r\ll\gamma^{*}$, $g^{(2)}_{ea}(\tau)$ is maintained once it rises from 0 to 1. However, when $\Omega_r\sim\gamma^{*}$  or $\Omega_r\gg\gamma^{*}$, the evolution of $g^{(2)}_{ea}(\tau)$ exhibits a damped oscillation described by $\Omega_r$  and $\gamma^{*}$ with the maximum amplitude exceeding 1, which is similar to the performance of a two-level system that is close to or in the strong excitation regime. This effect can be understood using the effective decay described in Sec.~\ref{Sec-2b}. The $\Lambda$-shape system considered here can be regarded as an effective two-level system shown in Fig.~\ref{Fig-1}(b), which is driven by a coherent field with the Rabi frequency $\Omega_r$ and the effective decay rate is $\gamma^{*}$. As stated in Sec.~\ref{Sec-2b}, the decay $|e\rangle\rightarrow |a\rangle$ inherits the property of the effective decay $|g\rangle\rightarrow |a\rangle$.  After a photon is emitted from the effective decay $|g\rangle\rightarrow |a\rangle$, this effective transition will be re-excited to the state $|e\rangle$ by F2 at the rate $\Omega_r$. When $\Omega_r\sim\gamma^{*}$  or $\Omega_r\gg\gamma^{*}$, the re-excitation rate of the state $|g\rangle$ approaches or exceeds the decay rate $\gamma^{*}$ at which the photon is emitted by the effective decay $|g\rangle\rightarrow |a\rangle$, so the emitted photons, which is essentially the photons from the decay $|e\rangle\rightarrow |a\rangle$, exhibit a clustered temporal distribution, i.e., $g^{(2)}_{ea}(\tau)>1$, as shown in Fig.~\ref{fig-3LS_g2tNS}(b). Moreover, the second-order correlation $g^{(2)}_{ea}(\tau)$ of the emitter’s transition is linked to the two-photon correlation $g^{(2)}_{s}(0)$ on the detection setup through convolution with the response function. Therefore, when the detection response time is in a proper region, this temporal effect of photons is detected, resulting in a bunching effect in the detection setup as shown in Fig.~\ref{Fig-3}(b).

\section{Effects of non-resonant pumping on single photons}
\label{App-detunning}
\begin{figure*}[ht]
	\centering\includegraphics[draft=false, width=1.35\columnwidth]{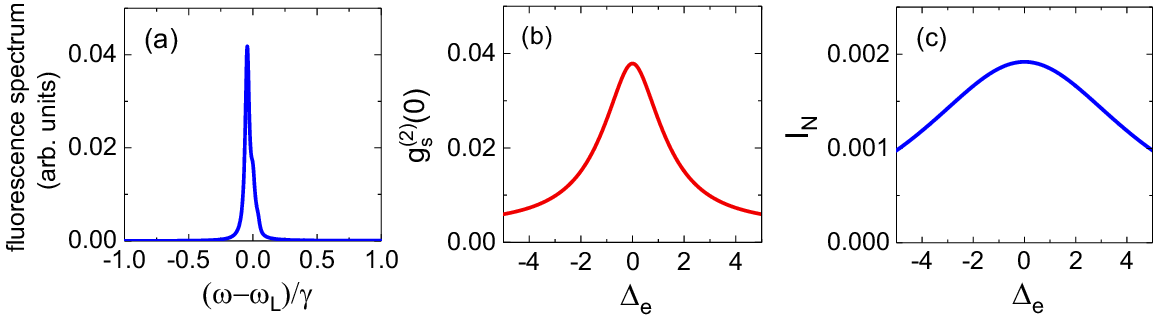}
	\caption{Fluorescence properties and detection response of the emission from the transition $|e\rangle\rightarrow |a\rangle$  with a detuning $\Delta_e$ between the transition $|g\rangle\leftrightarrow |e\rangle$  and the laser F1. (a) Fluorescence spectrum with $\Delta_e=2\gamma$ (b) Two-photon correlation $g^{(2)}_{s}(0)$ and (c) emission rate $I_{N}$ of the single photons as a function of $\Delta_e$. The detection bandwidth satisfies $\kappa=0.5\gamma $ in (b), and other parameters in (a)-(c) are the same as in Fig.~\ref{Fig-2}(a).  
\label{fig-Detuning}}
\end{figure*}

In the main text, we considered the case where the laser F1 resonantly drives the transition $|g\rangle\leftrightarrow |e\rangle$, i.e., $\Delta_e=0$.  In order to understand the effect of laser detuning on subnatural-linewidth single photons, we next investigate the case of $\Delta_e\neq 0$. Therefore, the Hamiltonian of the emitter in the motion equation~(\ref{Eq-a1}) is given by  
\begin{equation}
	\textit{H}_{A}= \Delta_e\sigma_{ee} + \Omega \sigma_{eg} + \Omega_r \sigma_{ga}  + \rm{H.c}.    
	\label{Eq-detun1}
\end{equation}
In the weak excitation regime of the emitter, i.e., $\Omega, \Omega_r\ll\gamma_1,\gamma_2$, the excited state can be adiabatically eliminated by the similar treatment described in Appendix~\ref{App-1}. Therefore, we obtain the motion equations containing only the degrees of freedom of the ground states as
\begin{equation}
	\dot \rho _g =  - i[H'_T,\rho _g] + \gamma _{1d}^*{\cal D}[\sigma _{gg}{\rho _g}] + \gamma _{2d}^*{\cal D}[\sigma _{ag}{\rho _g}],
	\label{Eq-detun2}
\end{equation}
with
\begin{equation}
H'_T = \Delta _{eff}\sigma _{gg} + (\Omega _r\sigma _{ga} + H.c).
	\label{Eq-detun3}
\end{equation}
Here $\gamma _{1d}^*{\cal D}[\sigma _{gg}{\rho _g}]$ and $\gamma _{2d}^*{\cal D}[\sigma _{ag}{\rho _g}]$ denotes the effective decays obtained from the adiabatic elimination of the excited state. The effective detuning and decay rates are, respectively, given by
\begin{align}
{\Delta _{eff}} &=  - \frac{{4{\Delta _e}{{\left| \Omega  \right|}^2}}}{{{{({\gamma _1} + {\gamma _2})}^2} + 4\Delta _e^2}},   \nonumber    \\
\gamma _{1d}^* &= \frac{{4{\gamma _1}{{\left| \Omega  \right|}^2}}}{{{{({\gamma _1} + {\gamma _2})}^2} + 4\Delta _e^2}},    \nonumber    \\
\gamma _{2d}^* &= \frac{{4{\gamma _2}{{\left| \Omega  \right|}^2}}}{{{{({\gamma _1} + {\gamma _2})}^2} + 4\Delta _e^2}}, 
\label{Eq-detun4}
\end{align}
which are reduced to 
\begin{align}
{\Delta _{eff}} &=  - \frac{{{\Delta _e}{{\left| \Omega  \right|}^2}}}{{{\gamma ^2} + \Delta _e^2}},   \nonumber    \\
\gamma _d^* = \gamma _{1d}^* &= \gamma _{2d}^* = \frac{{{{\left| \Omega  \right|}^2}}}{{{\gamma ^2} + \Delta _e^2}},
\label{Eq-detun5}
\end{align}
with $\gamma_1=\gamma_2=\gamma$. Therefore, the original $\Lambda$-shape system can be regarded as an effective two-level system driven by the coherent field F2 at the rate $\Omega_{r}$ with a frequency shift $\Delta _{eff}$ and effective decay rate $\gamma_d^*$. From Eq.~(\ref{Eq-detun5}), we see that $\gamma_d^*$ decreases with the increase of the detuning $\Delta _{e}$.

We show in Fig.~\ref{fig-Detuning}(a) the fluorescence spectrum of the transition $|e\rangle\rightarrow |a\rangle$ with a significant laser detuning $\Delta _e=2\gamma$. It reveals that the position of the peak in the fluorescence spectrum has a shift, which originates from the effective detuning  $\Delta_{eff}$ and is much smaller than the natural linewidth $\gamma$ of the transition $|e\rangle\rightarrow |a\rangle$. Besides, the full width at half maximum (FWHM) of the fluorescence spectrum decreases compared to the case of $\Delta _e=0$  shown in Fig.~\ref{Fig-2}(a), because $\Delta _e$ causes the reduction of the effective linewidth  $\gamma _d^*$ according to Eq.~(\ref{Eq-detun5}). Therefore, the linewidth of the single photons from the transition $|e\rangle\rightarrow |a\rangle$ can have a narrower linewidth than that at $\Delta _e=0$. Consequently, in a detector with a certain bandwidth, the two-photon response $g^{(2)}_{s}(0)$ at $\Delta_e\neq 0$ can be smaller than that at $\Delta_e= 0$ as shown in Fig.~\ref{fig-Detuning}(b). In addition, we see from Fig.~\ref{fig-Detuning}(c) that the emission rate of the single photons decreases with the increase of $\Delta _e$, but the decrease is slow. In summary, we can conclude that the laser detuning does not significantly degrade the quality of the subnatural-linewidth single photons.  

\section{Generalization and experimental feasibility }  
\label{App-5}

According to the discussion in the main text, we know that various emitters with $\Lambda$ or $\Lambda$-like structures can be used to generate subnatural-linewidth single photons.
For instance, as a system extensively investigated and employed in quantum technologies ~\cite{Volz2006_Observation, Hofmann2012_Heralded, Leent2020_Long-Distance}, the single ${}^{87}$Rb atom loaded into an optical dipole trap is suitable for the implementation of the proposed approach because the closed hyperfine transition $F_{g}=1\rightarrow F_{e}=0$ of the $D_{2}$ line of ${}^{87}$Rb has a $\Lambda$-like structure.  
Besides, our approach can be implemented based on the $\Lambda$-shape structure contained in the charged GaAs and InGaAs quantum dots in Voigt magnetic field~\cite{Dutt2005_Stimulated, He2015_Dynamically}, where a transition between an electron-spin ground state ($|\uparrow\rangle$ or $|\downarrow\rangle$) and trion excited state  ($|\downarrow\uparrow\Uparrow\rangle$ or $|\downarrow\uparrow\Downarrow\rangle$) can be driven by a laser resonantly, and the coherent transition between electron-spin ground states can be realized by a magnetic field or a stimulated Raman process.

\subsection{Transition $F_{g}=1\rightarrow F_{e}=0$ of $D_{2}$ Line of ${}^{87}$Rb}

\begin{figure}[ht]
\centering\includegraphics[draft=false, width=0.8\columnwidth]{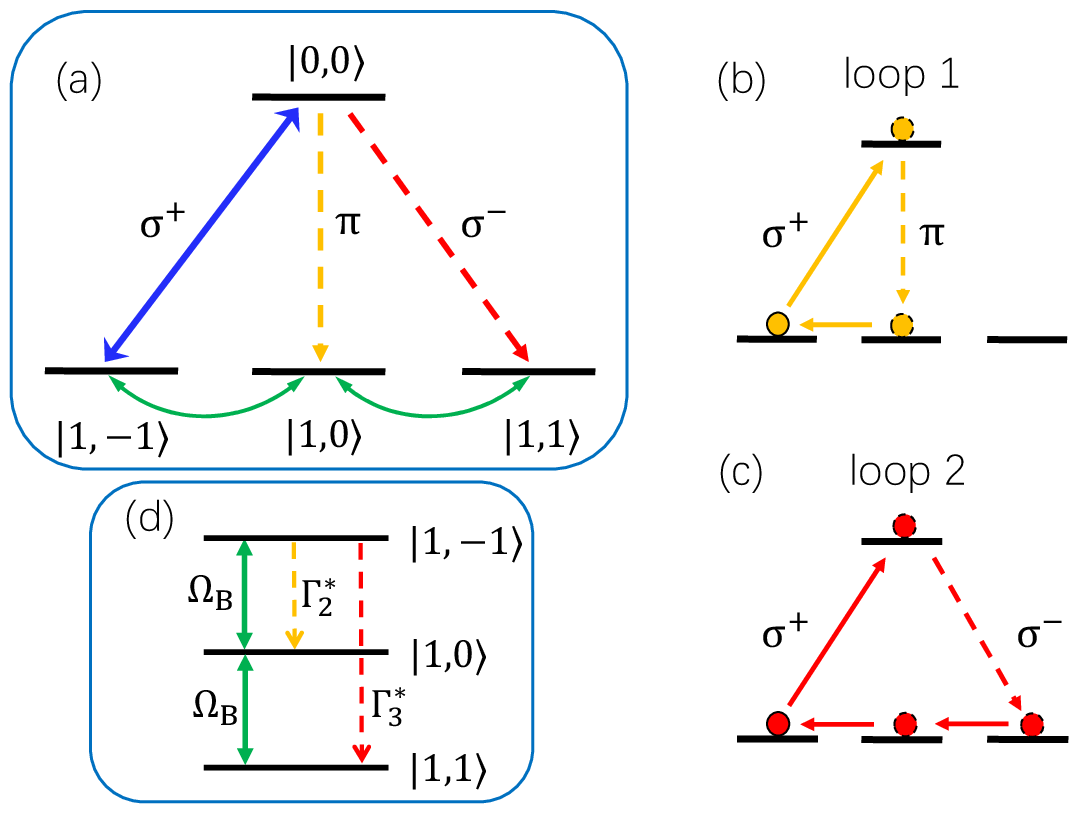}
\caption{Schematic diagram of generating subnatural-linewidth single photons based on the transition $F_{g} = 1 \rightarrow F_{e} = 0 $ of the $ D_{2}$ line of ${}^{87}$Rb. (a) The level configuration for the transition $F_{g} = 1 \rightarrow F_{e} = 0 $, where the transitions from the ground and excited states and between the ground-state Zeeman sublevels are, respectively, driven by $\sigma^{+}$-polarized laser field ($\textbf{E}$) and transverse magnetic field ($\textbf{B}_{T}$). Two main long-period transition loops involved in (a) are exhibited in (b) and (c), which dominate the emission from the spontaneous decays $|0,0\rangle\rightarrow|1,0\rangle$ and $|0,0\rangle\rightarrow|1,1\rangle$, respectively. (d) Schematic diagram of the effective three-level system of the emitter in (a) in the weak excitation regime.\label{fig-Rb87_level}}
\end{figure}

We choose the transition $F_{g}=1\rightarrow F_{e}=0$ of the $D_{2}$ line of ${}^{87}$Rb as the emitter to demonstrate the high experimental feasibility and generalizability of our study.  As shown in Fig.~\ref{fig-Rb87_level}(a), the ground hyperfine state $F_{g} =1$ is composed of three Zeeman sublevels $|1,-1\rangle, |1,0\rangle$, and $|1,1\rangle$, and the excited hyperfine state $F_{e} =0$ is composed of one Zeeman sublevel $|0,0\rangle$.
The transition $|1,-1\rangle\leftrightarrow|0,0\rangle$ is driven by a $\sigma^{+}$-polarized laser field $\textbf{E}$, and we set the quantization axis (z-axis) along the direction propagation of the light beam.
A static magnetic field $\textbf{B}_{T}$ perpendicular to the direction of light propagation is applied to drive the transitions between the ground-state Zeeman sublevels, whose direction is set along the x-axis.
Besides the above magneto-optical system, the transitions between the ground-state Zeeman sublevels can also be replaced by an optical scheme, i.e., stimulated Raman processes (not shown here).
The fluorescence emitted from the excited state to the ground state is collected by a detector, which is modeled by a quantized harmonic oscillator with bosonic annihilation operator $s$. In the frame rotating at the laser frequency $\omega_L$ and the rotating wave approximation, the time evolution of the combined system composed of the emitter and the filter is governed by the master equation
\begin{equation}
	\dot{\rho_r} =-\mathrm{i}[\textit{H}_{Rb},\rho_r]+ \sum_{i=-1}^{1} Y_i \mathcal{D} [\left|1,i\right\rangle\left\langle 0,0\right|] \rho_r  + \kappa \mathcal{D}[s] {\rho_r}.
	\label{Eq-V}
\end{equation}
Here $\rm{Y}_i=\rm{Y}/3$ are the decay rates from excited state $|0,0\rangle$ to ground state $|1,i\rangle$ with $i=0,\pm 1$, and $\rm{Y}$ is the natural linewidth  of the transition $F_{g}=1\rightarrow F_{e}=0$ of the $D_{2}$ line of ${}^{87}$Rb.
The Hamiltonian of the combined system is given by
\begin{equation}
	\textit{H}_{Rb}=\textit{H}_{0} + \textit{H}_{I} + \textit{H}_{B} + \textit{H}_{S}.     \label{Eq-13}
\end{equation}
$\textit{H}_{0}$ is the nonperturbed Hamiltonian of the emitter
\begin{equation}
	\textit{H}_{0}=\Delta_e |F_{e},m_{e}\rangle \langle F_{e},m_{e}|+H.c ,            \label{Eq-14}
\end{equation}
where $\Delta_e$ is the detuning of the transition $|1,-1\rangle\leftrightarrow|0,0\rangle$ frequency from the laser.
$H_I$ represents the laser-atom interaction Hamiltonian
\begin{equation}
\textit{H}_{I}=\sum_{g_{i}} V_{eg_{i}} |F_{e},m_{e}\rangle \langle F_{g},m_{g_{i}}|+H.c .        \label{Eq-15}
\end{equation}
According to the Wigner-Eckart theorem~\cite{Woodgate2000_Elementary, Brink1994_Angular, Meunier1987_A-simple}, the interaction energy for the transition $|F_{g},m_{g_i}\rangle\rightarrow|F_{e},m_{e}\rangle$ can be given by
\begin{align}
	V_{eg_{i}} &= -\langle F_{e},m_{e}|\textbf{d}|F_{g},m_{g_{i}}\rangle\cdot\textbf{E}    \nonumber    \\
	&= (-1)^{F_{e}-m_{e}+1}\left(\begin{array}{ccc}F_{e} & 1  & F_{g} \\ -m_{e} & q & m_{g_{i}} \\
	\end{array}\right)\Omega_{L} ,              
\label{Eq-16}
\end{align}
where $\textbf{d}$ is the electric dipole operator, $ \Omega_{L} = \left\langle F_{e}\parallel \textbf{d}\parallel F_{g}\right\rangle E $ is the Rabi frequency of the light field, and $q=0, \pm1$ denotes the spherical components.
$H_{B}$ represents the magnetic-atom interaction Hamiltonian
\begin{align}
	\textit{H}_{B} &= -\mu_{B}g_{F}\textbf{F}\cdot \textbf{B}_{T}     \nonumber    \\
	&= \mu_{B}g_{F}B_{T}\sum_{g_{i},g_{j}} \langle F_{g},m_{g_{i}}|\textbf{F}|F_{g},m_{g_{j}}\rangle \cdot\textbf{e}_{x} |F_{g},m_{g_{i}}\rangle \langle F_{g},m_{g_{j}}| ,        
\label{Eq-17}
\end{align}
where $\mu_B$ and $g_F$ denote the Bohr magneton and the gyromagnetic factor of the ground states, respectively, and
\begin{equation}
	\langle F_{g},m_{g_{i}}|\textbf{F}|F_{g},m_{g_{j}}\rangle
	\!=\!\left\langle \!F\!\parallel \!\textbf{F}\!\parallel \!F\!\right\rangle\! (-1)^{F_{g}-m_{g_{i}}} \!\left(\begin{array}{ccc}F_{g} & 1  & F_{g} \\ -m_{g_{i}} & q & m_{g_{j}} \\
	\end{array}\right) .              
\label{Eq-18}
\end{equation}
$\textit{H}_{S}$ is the Hamiltonian describing the detector and its coupling with the emitter
\begin{equation}
	\textit{H}_{S} = \Delta_{s} s^{\dag}s + (g A_{d}s^{\dag} + H.c ),                
\label{Eq-19}
\end{equation}
where the operator $A_d$ is defined as $A_d\equiv \left|1,d\right\rangle \left\langle 0,0\right|$ ($d=0,\pm1$), and denotes the atomic transition coupled with the detector.

In the weak excitation regime, i.e., $V_{eg_{-1}},\Omega_{B}\ll\rm{Y}$, we can deduce that there exist two main long-period transition loops similar to that in Fig.~\ref{Fig-1}(a), that is, transition loop 1 shown in Fig.~\ref{fig-Rb87_level}(b) which dominates the emission from the spontaneous decay $|0,0\rangle\rightarrow|1,0\rangle$,
and transition loop 2 shown in Fig.~\ref{fig-Rb87_level}(c) which dominates the emission from the spontaneous decay $|0,0\rangle\rightarrow|1,1\rangle$.
In transition loops 1 and 2, because the magnetic field driving the transitions between the ground states is weak, the emitter can stay in the states $|1,1\rangle$ and $|1,0\rangle$ for a long time, and thus these two states are both the metastable states.

Similar to the case in Fig.~\ref{Fig-1}, effective decays between the ground states are constructed by adiabatically eliminating the excited state, i.e.,
\begin{align}
\mathcal{L}_{\rm{eff}}\rho_{r,g}=\sum_{i=-1}^{1}Y^{*}_{i}\mathcal{D} [\left|1,i\right\rangle\!\left\langle 1,-1\right|] \rho_{r,g},
\label{Eq-20}
\end{align}
where $\rm{Y}^{*}_{i} = 4\rm{Y}_{i} |V_{e g_{-1}}|^2/\rm{Y}^2$ is the effective linewidth of the decay from the states $|1,-1\rangle$ to $|1,i\rangle$.
Therefore, the emitter in Fig.~\ref{fig-Rb87_level}(a) can be regarded as an effective three-level system driven coherently by the magnetic field as shown in Fig.~\ref{fig-Rb87_level}(d).
And, the effective decays $|1,-1\rangle\rightarrow|1,0\rangle$ and  $|1,-1\rangle\rightarrow|1,1\rangle$ correspond to the decays involved in loops 1 and 2, respectively.

\subsection{Fluorescence Properties and Detection Response}

\begin{figure*}[ht]
\centering\includegraphics[draft=false, width=1.5\columnwidth]{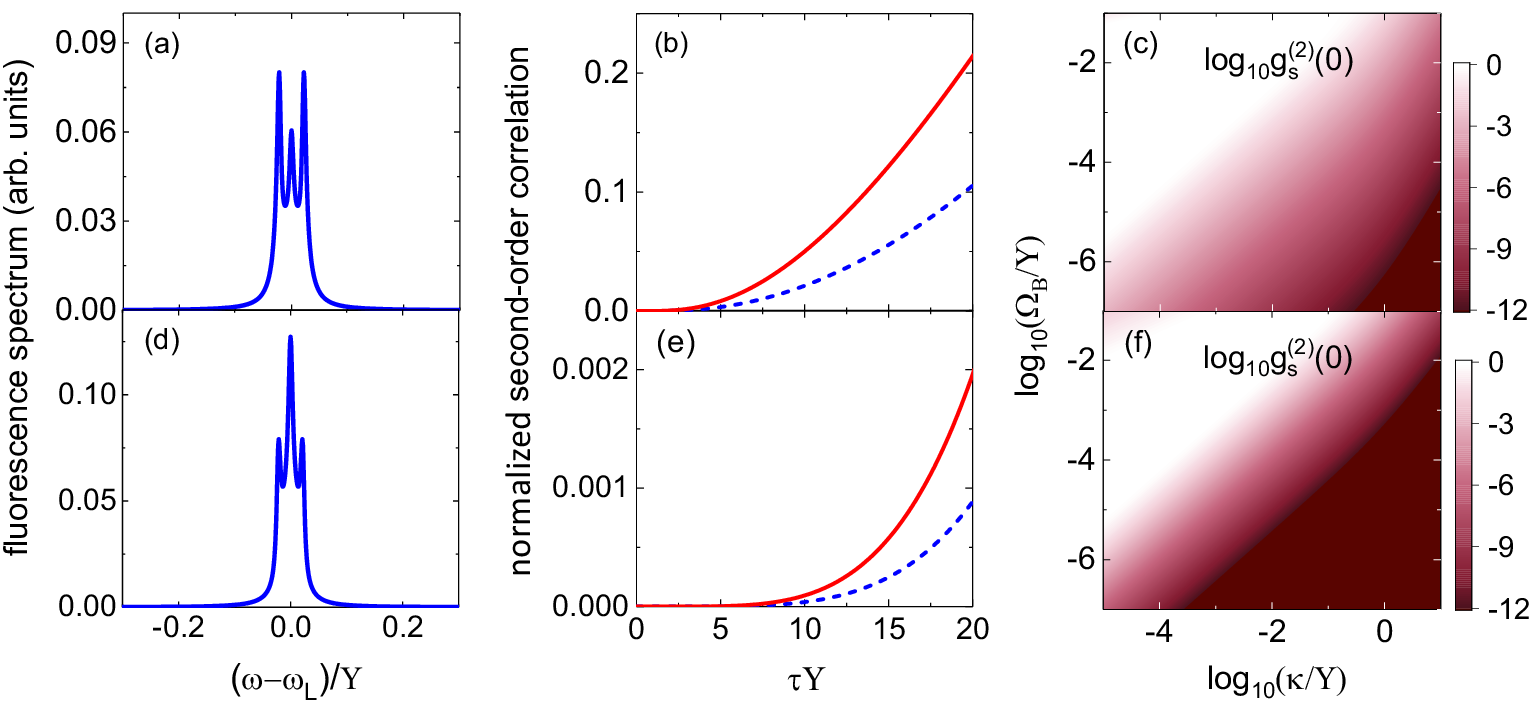}
\caption{Fluorescence properties and detection response of the emissiom from the transition $F_{g} = 1 \rightarrow F_{e} = 0 $ of the $ D_{2}$ line of ${}^{87}$Rb. The spectral property, second-ordering correlation, and detection response for the fluorescent photon emitted from the transitions $|0,0\rangle\rightarrow|1,0\rangle$ and $|0,0\rangle\rightarrow|1,1\rangle$ are, respectively, exhibited in (a)-(c) and (d)-(f). In (a) and (d), fluorescence spectrum for $\Omega_B=10^{-2}\rm{Y}$ and $V_{eg_{-1}}=10^{-1}\rm{Y}$. In (b) and (e), normalized second-ordering correlations as a function of delay $\tau$ for $\Omega_B=10^{-2}\rm{Y}$ and $V_{eg_{-1}}=10^{-1}\rm{Y}$ donated by red solid lines, and $\Omega_B=10^{-2}\rm{Y}$ and $V_{eg_{-1}}=10^{-2}\rm{Y}$ donated by blue dashed lines.  In (c) and (f), normalized two-photon correlations of the detector as a function of detection bandwidth $\kappa$ and the interaction energies of coherent fields with $\Omega_B=V_{eg_{-1}}$.\label{fig-Rb87_fluorescence}}
\end{figure*}

The fluorescence spectrum of the transitions $|0,0\rangle\rightarrow|1,0\rangle$ and $|0,0\rangle\rightarrow|1,1\rangle$ are shown in Figs.~\ref{fig-Rb87_fluorescence}(a) and (d).
Similar to the case in Fig.~\ref{Fig-2}(b), multi-peak structures arise, which can be understood by means of the effective level system in Fig.~\ref{fig-Rb87_level}(d). Taking transition $|0,0\rangle\rightarrow|1,0\rangle$ as an example, we explain the origin of the spectral structure.
As mentioned above, the effective decay $|1,-1\rangle\rightarrow|1,0\rangle$ shown in Fig.~\ref{fig-Rb87_level}(d) corresponds to the two-photon process composed of the transition $|1,-1\rangle\rightarrow|0,0\rangle$ driven by the laser and the subsequent spontaneous decay $|0,0\rangle\rightarrow|1,0\rangle$.
According to the standard dressed-state theory, there are four sidebands located at the frequencies $\pm\sqrt{2}\Omega_{B}$ and $\pm2\sqrt{2}\Omega_{B}$ in the emission of effective decay $|1,-1\rangle\rightarrow|1,0\rangle$~\cite{Zhang2019_Absorption}.
And since the laser frequency is constant, the transition $|0,0\rangle\rightarrow|1,0\rangle$ exhibits the same spectral structure as that in the effective decay $|1,-1\rangle\rightarrow|1,0\rangle$.
Therefore, one can infer that all spectral components of fluorescence are concentrated on a bandwidth of about $4\sqrt{2}\Omega_B$, which can be much smaller than the natural linewidth of the emitter.
From another perspective, we can conclude that because the emission from the spontaneous decays $|0,0\rangle\rightarrow|1,0\rangle$ and $|0,0\rangle\rightarrow|1,1\rangle$ are totally dominated by loops 1 and 2, respectively, the global spectral narrowing occurs.

In Figs.~\ref{fig-Rb87_fluorescence}(b) and (e), the statistical properties of fluorescence emitted from the transitions $|0,0\rangle\rightarrow|1,0\rangle$ and $|0,0\rangle\rightarrow|1,1\rangle$ are exhibited, respectively.
Because of the long periods of transition loops 1 and 2, the average delays between two successive fluorescent single photons emitted from the above transitions are long. Accordingly, the normalized second-order correlations of these two transitions can both remain a value very close to zero for a long delay which is much larger than the natural lifetime of the excited state.
Moreover, because loop 2 has a longer period than loop 1, the corresponding second-order correlation can be maintained near zero for a longer time.

According to the above discussion, we see that the nontrivial spectral and statistical properties similar to the $\Lambda$-shape system shown in Fig.~\ref{Fig-1}(a) can be well realized in the transitions $|0,0\rangle\rightarrow|1,0\rangle$ and $|0,0\rangle\rightarrow|1,1\rangle$ of transition $ F_{g}=1\rightarrow F_{e}=0$ of the $D_{2}$ line of ${}^{87}$Rb. Accordingly, the excellent single-photon responses on the detector with a bandwidth approaching or smaller than the natural linewidth of the emitter are predicted, which is confirmed in Figs.~\ref{fig-Rb87_fluorescence}(c) and (f).
In addition, because loop 2 has a longer period than loop 1, the single-photon response for the photons from the transition $|0,0\rangle\rightarrow|1,1\rangle$ on the detector is more conspicuous.

\section{Other realizable candidates}
\label{App-6}
\begin{figure}
\centering\includegraphics[draft=false, width=1\columnwidth]{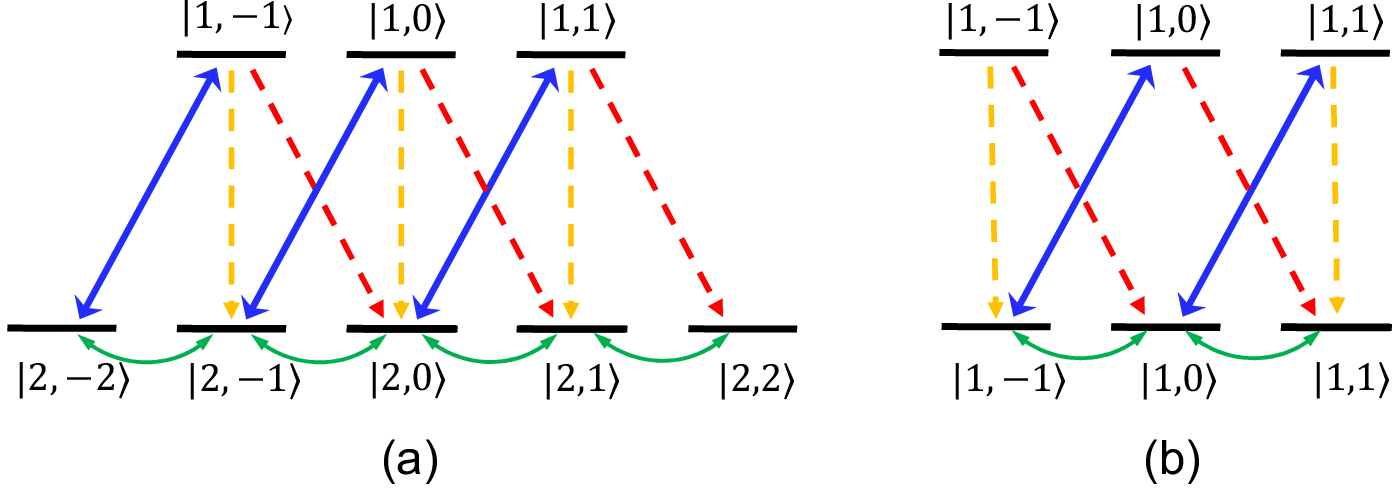}
\caption{Schematic diagrams of generating subnatural-linewidth single photons based on transitions (a) $F_{g}=2\rightarrow F_{e}=1$ (b) $F_{g}=1\rightarrow F_{e}=1$. The transitions between the ground states and the excited states are driven by a $\sigma^{+}$-polarized laser field, and photons with different polarizations are emitted by spontaneous decays. A transverse magnetic field is applied to drive the transitions between the ground-state Zeeman sublevels (the driving between the excited-state Zeeman sublevels is weak compared to the spontaneous emission process, and thus is ignorable). By choosing the fluorescent photon with a polarization different from the laser as the target photon, the single photons with subnatural linewidth are generated.\label{fig-otherLevel}}
\end{figure}

Besides the $\Lambda$-shape system and the closed transition $F_{g}=1\rightarrow F_{e}=0$ of the $D_{2}$ line of ${}^{87}$Rb, the approach for generating single photons with subnatural linewidth can also be implemented in many transition structures based on various alkali-metal atoms, e.g., in other kinds of transition $F_{g} \rightarrow F_{e}=F_{g}, F_{g}\pm1$.
Concretely, a circularly polarized laser drives the transitions between the ground and excited states, while a magnetic field perpendicular to the light propagation direction drives the transitions between the ground-state Zeeman sublevels, as shown in Fig.~\ref{fig-otherLevel}.
In the weak excitation regime of the laser and magnetic fields, the single photons with subnatural linewidth can be emitted from the transitions undriven by the laser due to the long periods of the transition loops.
In these schemes, the linewidth and emission rate of the single photons can be manipulated by two external coherent fields conveniently.

\begin{backmatter}

\bmsection{Funding}
National Key R\&D Program of China (Grant No. 2023YFA1407600);
National Natural Science Foundation of China (NSFC) (Grants No. 12275145, No. 92050110, No. 91736106, No. 11674390, No. 91836302, No. 11774118 and No. 11474119);
Fundamental Research Funds for the Central Universities of MOE (Grants No. CCNU18CXTD01 and No. CCNU17TS0006).

\bmsection{Acknowledgments}
He-bin Zhang would like to thank Jizhou Wu for his valuable discussions.

\bmsection{Disclosures}
The authors declare no conflicts of interest.

\bmsection{Data Availability}
All relevant data are available from the corresponding author upon request.

\end{backmatter}


\bibliography{bib_SNSP}

\end{document}